\documentclass[pra,floatfix,10pt]{revtex4}
\usepackage{color} 
\usepackage{epsfig}
\usepackage{amsmath}
\usepackage{graphicx}
\usepackage{subfig}

\begin{document}
\title
{ Photonic realization of the deformed Dirac equation via the segmented
graphene nanoribbons under inhomogeneous strain}
\author{M. R. Setare   $^{1}$,  P. Majari  $^{1}$,  C.  Noh   $^{2}$,   Sh. Dehdashti  $^{3,4}$ }

\affiliation{
$^1$ Department of Science, University of Kurdistan, Sanandaj, Iran\\
$^2$ Department of Physics, Kyungpook National University, Daegu 41566, Korea\\
$^3$School of Information Systems, Queensland University of Technology, Brisbane, Australia\\
$^4$Department of Electrical and Computer Engineering, University of Wisconsin – Madison, Madison, WI 53705, USA}


\begin{abstract}
Starting from an engineered periodic optical structure formed by waveguide arrays comprised of two interleaved lattices, we simulate a deformed Dirac equation. We show that the system also simulate graphene nano ribbons under strain. This optical analogue allows us to study the phenomenon of Zitterbewegung for the modified Dirac equation. Our results show that the amplitude of Zitterbewegung oscillations changes as the deformation parameter is changed.

\end{abstract}
\maketitle

\section{Introduction}
During the last decades, classical  analogues of quantum  and quantum-relativistic systems  have attracted much attention \cite{a0,a00,a000}. There have  been a significant movement towards simulating quantum phenomena in optical systems \cite{a1,a2,a3,a4,a5,a6,a7,a8,a9,a10}. These studies have eventually led to  a new class of analogue between classical optics and quantum mechanics, such as Aharonov-Bohm effect, quantum collapses and revivals and the Berry phase,~\textit{etc }\cite{a3,a4,a5}. Furthermore, a series of papers on classical simulation of relativistic quantum mechanics in the optical systems  have appeared including experimental realizations \cite{a7,a13,a16,a14,Keil,Koke}. One powerful method by which relativistic quantum phenomena are generated in the laboratory is photonic waveguides \cite{a016}.

 Graphene is an ideal candidate for simulating the Dirac equation in the lab. It can be imagined as a layer of carbon atoms  and its  electrons can be regarded as  relativistic particles.  Despite many  properties of graphene,  the absence of a band gap is the biggest obstacle  to be used in electronic devices \cite{g1,g2,g22}. In recent years, much effort has been made on engineering the band gap by applying strain. In this case, charge carriers obey a generalized Dirac equation\cite{g3,g4,g5}.

Photonic waveguides play a significant role in simulating relativistic quantum physics \cite{g6}. Indeed, one dimensional Dirac equation can be realized by spatial beam propagation in binary waveguide arrays  composed of two type of  equally spaced  waveguides \cite{a16}. It was shown that this setup can be described by a simplified system of coupled mode equations that allows us to obtain the standard Dirac equation under certain assumptions. In addition, photonic  waveguides can be engineered to simulate quantum phenomena \cite{a14,a014,a0014,a00014} and to realize the non-linear coherent states \cite{a0016,a316,a416}. It is important to emphasize that coherent states are extremely useful in physics. The concept of coherent states  has  been developed  in many different branches of physics such as mathematical physics \cite{a015} and quantum optics \cite{a0015}. Recently, the coherent states of deformed Heisenberg-Weyl algebra have been investigated  \cite{a15} by using a waveguide lattices with specific coupling coefficients between them.

 The motivations to study deformations are manifold\cite{a20,a21,b21}. One well-known deformation occurs  due to the so-called doubly special relativity which proposes that both the velocity of light, the Planck energy are universal constants \cite{a22}. The deformation can also emerge due to the existence of a minimum measurable length  \cite{a022}. Another famous deformations are caused by position-dependent mass where effective mass depend on the position \cite{mm1,mm2,mm3,mm4}, which is used in semiconductor heterostructures \cite{mm5}, quantum dots \cite{mm6}, semiconductor theory \cite{mm7} and  problems in condensed-matter physics \cite{mm8}. Deformed canonical commutations are obtained  by modifying the metric  structure  associated with a curved space \cite{mm9,mm10, mm11}. In this approach there is a causal relationship between deforming function and the metric tensor \cite{mm12,mm13}. 


In this work, we propose a 1D periodic array of coupled waveguides in which separations between the waveguides are controlled in order to simulate deformed Dirac equation.
We first construct the deformed relativistic wave equation making use of deformed Lie algebras. This deformation plays the role of nonlinearities in our model. We then discuss Zitterbewegung (ZB), an extremely fast oscillation of relativistic particles. We compare the ZB effect in the deformed model against the original one, showing that the amplitude and  frequency change with the deformation parameter.
The paper is organized as follows. In Sec. II, we introduce segmented graphene nano ribbons under strain which provides an  experimental tool to realize a generalized Dirac equation.  We then show how the same model arises in engineered photonic waveguides in Sec. III,  and study the ZB in the deformed scenario in Sec. IV. We conclude in Sec. V.

\section{Segmented graphene nanoribbons under strain}

Graphene is a single atomic layer of carbon arranged in a honeycomb lattice.  Among the carbon nanostructures, graphene Nanoribbons(GNR) that are narrow strips of graphene have garnered great interest in recent years. As shown in Ref.~\cite{b24} for narrow GNRs we have the following 1D generalized Dirac equation:
\begin{equation}\label{aa1}
H \psi=-i \hbar v_f \alpha {\partial \psi \over \partial x}+\phi(x)\beta \psi
\end{equation}
where $\phi(x)$ is position dependent scalar Higgs field. Here we choose $\phi(x)\equiv \phi_0$, in which case the equation becomes the usual 1D Dirac equation \cite{b25}. The effects of nonisotropic strain on GNR can be summarised by the substitution $p_x\rightarrow f(x)p_x$ in the above Hamiltonian \cite{b26}:
\begin{equation}\label{aa2}
H \psi=-i \hbar v_f f(x)\alpha {\partial \psi \over \partial x}+\phi_0 \beta \psi
\end{equation}

\begin{figure}
  \centering
  \includegraphics[width=0.40\textwidth]{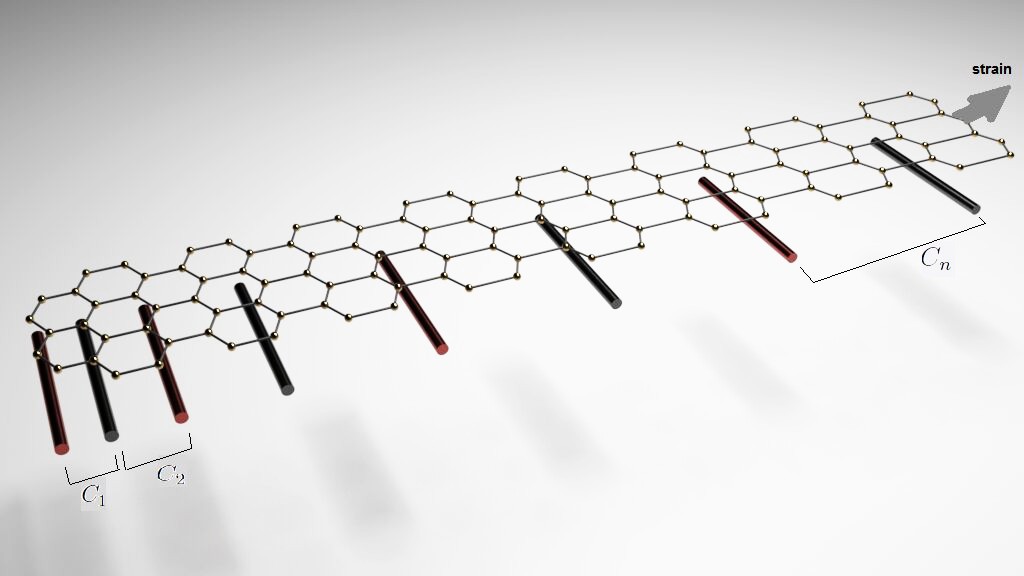}
  \caption{Schematic of  graphene nanoribbons under strain and a binary array  made of two   waveguides A and B   with black and red color respectively. }\label{sh1}
\end{figure}
We will show that this deformed Dirac equation can also be simulated by a specifically engineered waveguide array.

\section{Simulation of the deformed Dirac equation by binary photonic superlattices}

Within the nearest-neighbour coupling approximation the propagation of an optical field in disordered waveguide arrays, comprised of two interleaved lattices A and B, is described by the following equations \cite{0,1}:
 \begin{equation}\label{0}
-i{ d E_n \over dz}=(-1)^n \beta  E_n+ c _{n,n-1}E_{n-1} +c _{n,n+1}E_{n+1},
\end{equation}
where $E_n$ is the electric field amplitude at the nth waveguide ($n = 1,2,... ,N$), $\beta$ is the propagation constant and
$c _{n,n\pm1}$  are the coupling coefficients between waveguides. As we show in Fig.~({\ref{sh1}}) the coupling constant depends on the
separation distance \cite{001}. Thus we can control the coupling coefficients, by controlling the separations between waveguide elements.
Here we choose the coupling coefficients as: $c _{n,n-1}= c_1{f(n)}$  at any site $n>1$ \cite{a15}. Consequently, the coupled-mode equations take the following form \cite{2}:\begin{equation}\label{2}
i {d E_{2n}\over {dz}}+c_1 {f(2n+1)}E_{2n+1}+c_1 {f(2n)}E_{2n-1}-\beta E_{2n}=0,
\end{equation}
and
\begin{equation}\label{3}
i {d E_{2n-1}\over {dz}}+c_1 {f(2n)}E_{2n}+c_1 {f(2n-1)}E_{2n-2}+\beta E_{2n-1}=0.
\end{equation}
By setting $E_{2n}=(-1)^n \psi_1(2n,z)$, $E_{2n-1}=-i(-1)^n \psi_2(2n-1,z)$, Eqs.~({\ref{2}}) and ({\ref{3}}) become
\begin{equation}\label{6}
i {d \psi_1(2n,z)\over {dz}}+ic_1 {f(2n+1 )}\psi_2 (2n+1,z)-i c_1 {f(2n)}\psi_2(2n,z)-\beta \psi_1(2n,z)=0,
\end{equation}
and
\begin{equation}\label{7}
i {d \psi_2(2n,z)\over {dz}}+i c_1 {f(2n )}\psi_1 (2n,z)-i c_1 {f(2n-1)}\psi_1(2n-1,z)+\beta \psi_2(2n,z)=0.
\end{equation}

We consider the deformation \cite{4}
\begin{equation}\label{007}
f(n)=\sqrt{1+{\chi_a \mu  \over \nu}(1-n)},
\end{equation}
in which $\chi_a$ is an anharmonicity parameter ($0\leq \chi_a \leq \nu$ ) and $\mu=\pm1$. To achieve this in our system we choose  $c _{n,n-1}= c_1    e^{- d_{(n,n-1)}+\kappa d_1\over {d_1}}$, where $d_1$ is the distance
between the first coupled waveguides and  $\kappa$  is a positive constant. By choosing  ${\chi_a \over \nu}={1\over 100}$, one can make an approximation $f(2n\pm 1)\simeq f(2n)$ which s necessary in order for our system to mimic the relativistic wave equation. Indeed, we can rewrite  Eqs.~({\ref{6}}) and ({\ref{7}}) as
\begin{equation}\label{8}
i {d \psi_1(2n,z)\over {dz}}+c_1{f(2n)}(i \psi_2 (2n+1,z)-i  \psi_2(2n,z))-\beta \psi_1(2n,z)=0,
\end{equation}
and
\begin{equation}\label{9}
i {d \psi_2(2n,z)\over {dz}}+c_1{f(2n)}(i  \psi_1 (2n,z)-i \psi_1(2n-1,z))+\beta \psi_2(2n,z)=0.
\end{equation}

Finally, by employing  $\psi(n\pm 1,z)=\psi(n,z)\pm {\partial \psi(n,z) \over \partial n}$, the  coupled-mode equations can be written in the following form:
 \begin{equation}\label{10}
i {d \psi_1(2n,z)\over {dz}}=-ic_1 f(2n) {\partial \psi_2(2n,z) \over \partial n}+\beta \psi_1(2n,z),
\end{equation}
and
\begin{equation}\label{11}
i {d \psi_2(2n,z)\over {dz}}=-ic_1 f(2n) {\partial \psi_1(2n,z) \over \partial n}-\beta \psi_2(2n,z).
\end{equation}

 Therefore, making a replacement $n\rightarrow x$  and substituting Eq.~({\ref {007}}) into Eqs.~({\ref {10}}) and ({\ref {11}}) yields
\begin{equation}\label{12}
i {d \psi\over {dz}}=c_1 f(x)p_{x} \sigma_x \psi+\beta \sigma_z \psi= c_1 P_{x} \sigma_x \psi+\beta \sigma_z \psi,
\end{equation}
where $\psi=(\psi_1(x,z), \psi_2(x,z))^\dag$ and $P_x = -i\partial_X=-if(x)\partial_x$ with $X=-{2  \mu \nu \over \chi_a}\sqrt{1+{\chi_a  \mu\over \nu}(1-x)}$. The above expression, after making formal changes $z \rightarrow t$, $\beta\rightarrow m$, $c_1\rightarrow c$ and setting $\hbar=c=1$ , reduces to the sought-for deformed Dirac equation.
Notice that we can interpret the above deformed momentum in terms of position-dependent-mass formalism. In this case, mass  is a  function of  the position as follows \cite{p0,p1}:
\begin{equation}\label{c12}
P_x={1 \over \sqrt{m(x)}}p_x.
\end{equation}
Another way to  interpret the deformed momentum in Eq.~(\ref{12}) is in terms of the  Dirac equation in a curved space  \cite{p2,p3,p4,p5,p6} with the metric \cite{p7}
\begin{equation}\label{cc12}
g(x)={1 \over f^2(x)}.
\end{equation}

\section{Zitterbewegung in the deformed model}
The trembling motion of relativistic particles caused by an interference between positive and negative energy wave components is known as Zitterbewegung  \cite{m4,m04,m004}. To simulate ZB in our waveguide setup the elements in the array are excited  by a broad beam $E(x,0)$ with  the propagating field envelopes  $G(X)$.    We  assume that $G(X)$ change slowly over the waveguide spacing, i.e., $G(c_1 f(2n))\simeq G(c_1 f(2n-1))$. The solution to  Eq.~({\ref {12}}) in the momentum space reads
\begin{figure}
\subfloat[]{\includegraphics[width=0.40\textwidth]{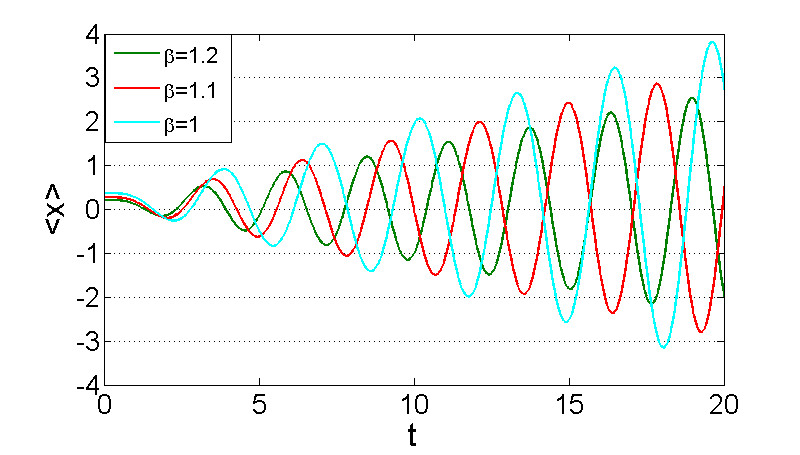}}
\hspace{1cm}
\subfloat[]{\includegraphics[width=0.40\textwidth]{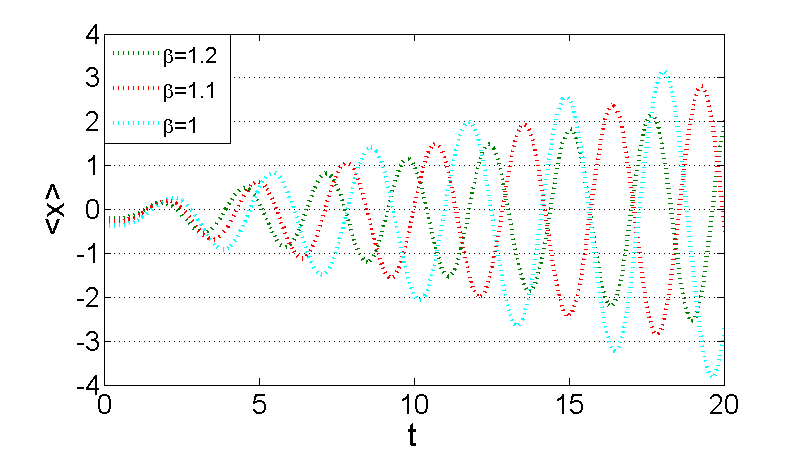}}
  \hfill
  \\
\subfloat[]{\includegraphics[width=0.40\textwidth]{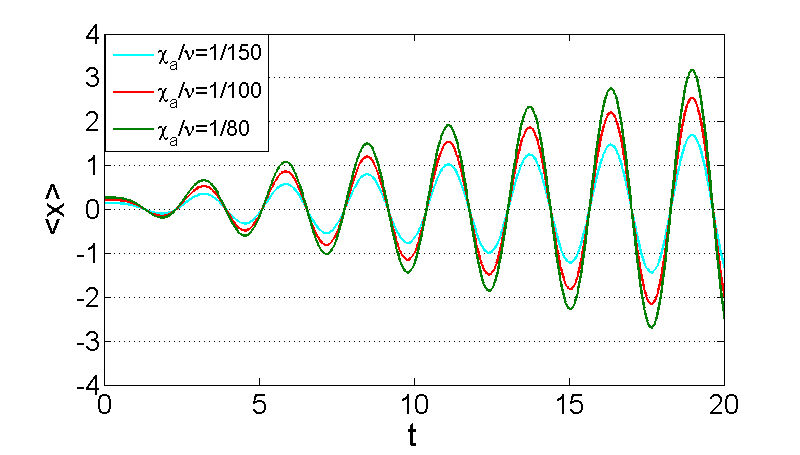}}
\hspace{1cm}
\subfloat[]{\includegraphics[width=0.40\textwidth]{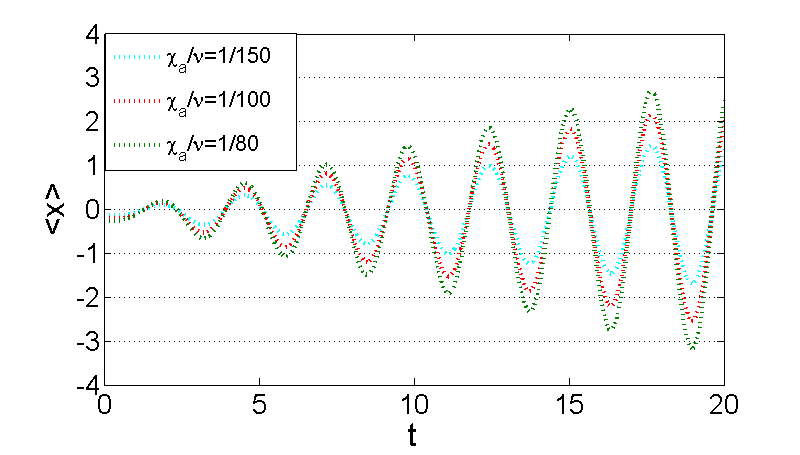}}
  \caption{Corresponding behavior of
 $<x>_t$  with $\mu=-1$(solid curve) and $\mu=1$  (dashed curve) for $G(x)\propto{1\over 9+x^2}$(a,b) as a function of
the $\beta$ with ${\chi_a \over \nu}={1\over 100}$(c,d) as a function of
the ${\chi_a \over \nu}$ with $\beta=1.2$.}\label{sh2}
\end{figure}
\begin{equation}\label{14}
\psi_1(k,z)=G(k)[cos( \varepsilon z )-i(c_1 k+\beta ){sin(\varepsilon z)\over \varepsilon}],
\end{equation}
and
\begin{equation}\label{15}
\psi_2(k,z)=G(k)[cos( \varepsilon z )+i(-c_1 k+\beta ){sin(\varepsilon z)\over \varepsilon}],
\end{equation}
where   $\varepsilon=\sqrt{{c_1}^2 k^2+{\beta}^2}$ and $G(k)={1\over 2\pi}\int dX G(X)e^{-ikX}$.  After some straightforward calculations, the average   position in the array  becomes
\begin{multline}\label{16}
<x>_z=<x>_0+{{\mu \pi  \chi_a \over \nu}}\int{G^*(k)\partial^2 G(k)dk}-{\mu \pi {c_1}^2 \chi_a \over 2\nu}\bigg\{\int{ 2|G(k)|^2 ({(\varepsilon ^2 z^2-2)\over \varepsilon ^3}) dk}+\\
\int{ 2|G(k)|^2 ({(\varepsilon ^2 z+ikc_1)\over \varepsilon ^4}) sin(2\varepsilon z) dk}
-\int{ |G(k)|^2 ({4(ikzc_1-1)\over \varepsilon ^3}) cos^2(\varepsilon z) dk}\bigg\}. \hspace{12.0em}
\end{multline}

The last two terms in the above equation are oscillatory terms that yields ZB as shown in Fig.~({\ref{sh2}}). The amplitude of oscillation changes because of  the different periodicity in the x direction($\mu=1$ and $\mu=-1$ which means as n increase  the separation distance  between  waveguides  decrease and increase respectively).

In our  system $\beta$ is the propagation constant and  plays the role of mass in the deformed Dirac equation. Figures 2(a) and (b) show that the amplitude of oscillation decreases while the frequency increases as the mass is increased. Changes in ZB are different as one varies the deformation parameter as depicted in Figs.~2(c) and (d). The amplitude of ZB increases while the frequency stays the same, as the deformation parameter is increased. Note that, in the limit ${\chi_a \over \nu}\rightarrow0$, one gets back the usual model which there is not any change in the  amplitude of oscillations.

\section{Conclusions}

We have derived a coupled  equations for light propagation in the waveguide lattices which is comprised of two interleaved lattices with specifically engineered coupling coefficients between neighbouring waveguides. We showed that the resulting equation is equivalent to a deformed Dirac equation that arises in graphene nano ribbons under strain and also showed a connection between the deformed Dirac equation with position-dependent mass and curved space. Lastly, we have calculated the average position in the deformed model and found that ZB changes as the deformation parameter is changed. 

\end{document}